\documentclass[12pt,thmsa]{article}
\usepackage{amssymb}

\usepackage{amsmath}

\input{tcilatex}

\begin{document}

\author{\textbf{Chong Li}\thanks{%
lichong@student.dlut.edu.cn}\textbf{\ },\textbf{\ He-Shan Song, and Ling Zhou%
} \\
Department of physics,\\
Dalian University of Technology, \\
Dalian City, 116023 P.R.China}
\title{\textbf{Alternative new notation for quantum information theory}}
\date{May}
\maketitle

\begin{abstract}
A new notation has been introduced for the quantum information theory. By
this notation,some calculations became simple in quantum information theory
such as quantum swapping, quantum teleportation.
\end{abstract}

\section{Introduction}

Since the concept of entanglement[$1$] was introduced into quantum physics,
it plays an important roles in quantum information theory. Nearly all of
quantum information processes are based on entangled state such as quantum
code[$2$], quantum swapping[$3$], quantum teleportation[$4$] and so on.

In the quantum information theory some calculations are very complex. For
example, in quantum teleportation\ we must write out all the joint states,
and need to find out the relation between the \emph{Alice}'s joint \emph{Bell%
} states measured results and the receiver \emph{Bob}'s local operation for
his particle. Furthermore, if we perform a teleptortation of $n$-dimension
state, the calculations are tremendous! Evenmore we make a series of $2$%
-dimensional quantum entanglement swapping, the calculations are very
difficult.

We have introduced a kind of notation[$11$], which was represented in Ref.[$%
10$], for quantum teleportation. In that paper, we give a powerful theorem
for general teleportation, by which one can easily construct \emph{Bob}'s
operation.

In this paper we generalize the notation to much more quantum information
process. By using this tool,we give a new criteria for \emph{Schmidt}
decomposition of entangled state. In terms of this criteria we need not to
calculate partial inner-product for the joint density operator.

This notation can simplify calculation of quantum swapping include series
quantum swapping,

As we know, notation is often used in physics, it can simplify some
calculation just as Dirac sign does. Here we try to make following notation
for quantum information.

In the paper[$11$],we introduced a notation for quantum teleportation,which
is defined as follows,

Let $\mathbf{h}$ be a finite \emph{Hilbert} space, and $\dim (\mathbf{h})=N$%
, so $2$-partite \emph{Hilbert} space $\mathbf{H}=h\otimes h$.

We definite some useful notations $:$

\begin{definition}
$\mathbf{\alpha }=\left( 
\begin{array}{cccc}
\alpha _0 & \cdots & \alpha _{N-2} & \alpha _{N-1}%
\end{array}
\right) $,
\end{definition}

\begin{definition}
$\mathbf{e}_i=\left( 
\begin{array}{cccc}
\left| 0\right\rangle _i & \cdots & \left| N-2\right\rangle _i & \left|
N-1\right\rangle _i%
\end{array}
\right) $,
\end{definition}

\begin{definition}
$\mathbf{\tilde{e}}_i=\left( 
\begin{array}{cccc}
_i\left\langle 0\right| & \cdots & _i\left\langle N-2\right| & 
_i\left\langle N-1\right|%
\end{array}
\right) $.
\end{definition}

Let set $\left\{ 
\begin{array}{cccc}
\left| 0\right\rangle _{i} & \cdots & \left| N-2\right\rangle _{i} & \left|
N-1\right\rangle _{i}%
\end{array}%
\right\} $ be a group of orthonormal complete basis of the $i$-th \emph{%
Hilbert} space $\mathbf{h}_{i}$. So an arbitrary state $\left| \varphi
\right\rangle =\overset{N-1}{\underset{i=0}{\sum }}\alpha _{i}\left|
i\right\rangle $ can be written as $\left| \varphi \right\rangle =\mathbf{%
\alpha e}^{t}$ (where superscrip $t$ means transposition). Arbitrary
bipartite state $\left| \psi _{12}\right\rangle =\underset{ij}{\sum }%
a_{ij}\left| i\right\rangle _{1}\left| j\right\rangle _{2}$ can be written
as $\left| \psi _{12}\right\rangle =$ $\mathbf{e}_{1}\mathbf{Ae}_{2}^{t}=%
\mathbf{e}_{2}\mathbf{A}^{t}\mathbf{e}_{1}^{t}$, where $\mathbf{A}$ is a
known matrix $\mathbf{A}_{ij}=a_{ij}$ and $A^{t}$ is the transposed of $A$.
If we perform a local operation $U_{1}$ on particle $1$, the state become $%
U_{1}\left| \psi _{12}\right\rangle =$ $\mathbf{e}_{1}U_{1}\mathbf{Ae}%
_{2}^{t}$, and it is $U_{2}\left| \psi _{12}\right\rangle =$ $\mathbf{e}_{1}%
\mathbf{A}U_{2}^{t}\mathbf{e}_{2}^{t}$ with local operation $U_{2}$.

We will find the merits of this notation from the following discussion.

\section{Application of the notation for entanglement research}

The phenomenon of entanglement is a remarkable feature of quantum theory. It
play a crucial role in the discussions of quantum mechanics and quantum
information theory. The entangled states are important for quantum
information theory. In particular, the pure bipartite entangled state $%
\left( \text{PBES}\right) $ is used in most quantum information transported
process such as quantum teleportation, quantum code et al. So how to find
out the PBES from PBS(pure bipartite state) is important. As we know, any
pure bipartite entangled state can be expressed as \emph{Schmidt}
decomposition, so the criteria for pure bipartite entangled state is, if the 
\emph{Schmidt} number of the pure bipartite state is more than one. by the
notation, a pure bipartite state $\left| \psi _{12}\right\rangle $ can be
described as

\begin{eqnarray*}
\left| \psi _{12}\right\rangle &=&\underset{ij}{\sum }a_{ij}\left|
i\right\rangle _{1}\left| j\right\rangle _{2} \\
&=&\left( \left| 0\right\rangle _{1}\text{ }\left| 1\right\rangle _{1}\cdots
\left| n-1\right\rangle _{1}\right) A\left( \left| 0\right\rangle _{2}\text{ 
}\left| 1\right\rangle _{2}\cdots \left| n-1\right\rangle _{2}\right) ^{t} \\
&=&\mathbf{e}_{1}\mathbf{Ae}_{2}^{t}.
\end{eqnarray*}

Where all entangled information are included in the matrix $A$. Here we
proposed a criteria for this.

$\emph{Criteria}$ $1:$ \textit{The state }$\left| \psi _{12}\right\rangle $%
\textit{\ is entangled state, if and only if there are more than one
non-zero eigenvalues of the matrix }$AA^{\dagger }$,\textit{the number of
the non-zero eigenvalues \textit{is} just} \emph{Schmidt} \textit{number}.

\emph{Proof}$:$As we know, any bipartite entangled state can be written as 
\emph{Schmidt} decomposition, namely

\begin{equation*}
\left| \psi _{12}\right\rangle =\underset{ij}{\sum }a_{ij}\left|
i\right\rangle _1\left| j\right\rangle _2=\underset{k}{\sum }\sqrt{P_k}%
\left| \widetilde{k}\right\rangle _1\left| \widetilde{k}\right\rangle _2
\end{equation*}

the joint density operator is

\begin{equation*}
\rho \left( \psi _{12}\right) =\underset{ijlm}{\sum }a_{ij}\left(
a_{lm}\right) ^{\dagger }\left| i\right\rangle _1\left| j\right\rangle
_2\left\langle m\right| _2\left\langle l\right| _1
\end{equation*}

and the density operator of the qubit $1$ is

\begin{eqnarray*}
\rho _1\left( \psi _{12}\right) &=&\underset{ilm}{\sum }a_{im}\left(
a_{lm}\right) ^{\dagger }\left| i\right\rangle _1\left\langle l\right| _1 \\
&=&AA^{\dagger }=B
\end{eqnarray*}

We can calculate out the eigenvalues of $\rho _{1}\left( \psi _{12}\right) $

\begin{equation*}
\det \left( \lambda -B\right) =0\Longrightarrow \mathbf{\lambda }
\end{equation*}

where $\mathbf{\lambda =}\left( \lambda _{1}\lambda _{2}\cdots \lambda
_{n}\right) $ and \emph{Schmidt} number $N_{s}\geqslant 2$ then the state $%
\left| \psi _{12}\right\rangle $ is entangled state, namely there are more
than two non-zero eigenvalue of the state $\left| \psi _{12}\right\rangle $.

\begin{equation*}
\left| \widetilde{k}\right\rangle _{1}=\underset{j}{\sum }\acute{t}%
_{kj}\left| j\right\rangle _{1}\text{ and }\rho _{1}\left( \psi _{12}\right)
\left| \widetilde{k}\right\rangle _{1}=\lambda _{k}\left| \widetilde{k}%
\right\rangle _{1}
\end{equation*}

Let $\left| \psi _{12}\right\rangle =\underset{k}{\sum }\sqrt{\lambda _{k}}%
\left| \widetilde{k}\right\rangle _{1}\left| \widetilde{k}\right\rangle
_{2}, $where $\left| \widetilde{k}\right\rangle _{2}=\underset{i}{\sum }%
\alpha _{i}^{k}\left| i\right\rangle _{2}$ with $_{2}\left\langle \widetilde{%
k}\right. \left| \widetilde{j}\right\rangle _{2}=\delta _{kj}$, considered $%
\left| \psi _{12}\right\rangle =\underset{ij}{\sum }a_{ij}\left|
i\right\rangle _{1}\left| j\right\rangle _{2}$, we can get all $\alpha
_{i}^{k}$. So $\left| \widetilde{k}\right\rangle _{2}$ is known. $\left|
\psi _{12}\right\rangle =\underset{k}{\sum }\lambda _{k}\left| \widetilde{k}%
\right\rangle _{1}\left| \widetilde{k}\right\rangle _{2}$ $\blacksquare $

From this criteria we not only can get the \emph{Schmidt} number of the
state, but also can find out the \emph{Schmidt }decomposition expression for
the state.

$\emph{Criteria}$ $2:$ \textit{The state }$\left| \psi _{12}\right\rangle $%
\textit{\ is entangled state, if and only if the rank of the matrix }$A$ 
\textit{is more than one, it is just the} \emph{Schmidt} \textit{number}.

\emph{Proof}$:$As we know,the non-zero eigenvalue number of any matrix is
equal to its rank.

There must be matrix $T$ , $TT^{\dagger }=M_{ii}\delta _{ij}$ with $%
M_{ii}\in \left\{ 0,1\right\} $

\begin{equation*}
TAT^{\dagger }=K,\text{with }K_{ij}=a_{ij}\delta _{ij}
\end{equation*}

and

\begin{equation*}
TA^{\dagger }T^{\dagger }=K^{\dagger },\text{with }K_{ij}^{\dagger
}=b_{ij}\delta _{ij}
\end{equation*}

so

\begin{equation*}
T\rho _{1}T^{\dagger }=TAT^{\dagger }TA^{\dagger }T^{\dagger }=KK^{\dagger
}=\sigma _{ij}\delta _{ij}
\end{equation*}

with $\sigma _{kk}=a_{kk}b_{kk}$. finally we can easily get

\begin{equation*}
R\left( \rho _{1}\right) =R\left( A\right) =R\left( A^{\dagger }\right) .
\end{equation*}

$\blacksquare $

\section{Application for quantum swapping and bipartite state quantum
teleportation}

Quantum information attracted more attention since quantum teleportation has
been proposed. Quantum swapping is important in quantum information. But
there are some complicated calculation in quantum cryptography[$14$] and
quantum swapping, furthermore multi-dimension teleportation or series of
swapping, while by this notation, all of this kinds of calculation become
easy.

\subsection{Quantum swapping}

As we know, any quantum information process is based on entangled state. If
sender and receiver do not share any common entangled state, how they to
communicate each other? The answer is that they can make a quantum swapping
through another person as follow. \emph{Alice} (sender) has a entangled
states $\left| \phi _{12}\right\rangle $, \emph{Bob}(receiver) has a
entangled states $\left| \varphi _{34}\right\rangle $, \emph{Alice} send
qubit $2$ to Charley(the third person) and \emph{Bob} send a qubit $3$ to
him, then Charley make a joint measurement on qubit $2$ and $3$, he told 
\emph{Alice} and \emph{Bob} his measured result, then \emph{Alice} and \emph{%
Bob} share a new entangled state $\left| \psi _{14}\right\rangle $, this
process is quantum swapping.

\begin{equation*}
\left| \phi _{12}\right\rangle =\left( \left| 0\right\rangle _{1}\text{ }%
\left| 1\right\rangle _{1}\cdots \left| n-1\right\rangle _{1}\right) A\left(
\left| 0\right\rangle _{2}\text{ }\left| 1\right\rangle _{2}\cdots \left|
n-1\right\rangle _{2}\right) ^{t}
\end{equation*}

\begin{equation*}
\left| \varphi _{34}\right\rangle =\left( \left| 0\right\rangle _{3}\text{ }%
\left| 1\right\rangle _{3}\cdots \left| n-1\right\rangle _{3}\right) C\left(
\left| 0\right\rangle _{4}\text{ }\left| 1\right\rangle _{4}\cdots \left|
n-1\right\rangle _{4}\right) ^{t}
\end{equation*}

measured result

\begin{equation*}
\left| \phi _{23}^{^{\prime }}\right\rangle =\left( \left| 0\right\rangle
_{2}\text{ }\left| 1\right\rangle _{2}\cdots \left| n-1\right\rangle
_{2}\right) B\left( \left| 0\right\rangle _{3}\text{ }\left| 1\right\rangle
_{3}\cdots \left| n-1\right\rangle _{3}\right) ^{t}
\end{equation*}

so the new entangled state

\begin{eqnarray*}
\left| \psi _{14}\right\rangle &=&\left\langle \phi _{23}^{/}\right| \left(
\left| \phi _{12}\right\rangle \otimes \left| \varphi _{34}\right\rangle
\right) \\
&=&\left( \left| 0\right\rangle _{1}\text{ }\left| 1\right\rangle _{1}\cdots
\left| n-1\right\rangle _{1}\right) A\overline{B}C\left( \left|
0\right\rangle _{4}\text{ }\left| 1\right\rangle _{4}\cdots \left|
n-1\right\rangle _{4}\right) ^{t} \\
&=&\rho \left( \left| 0\right\rangle _{1}\text{ }\left| 1\right\rangle
_{1}\cdots \left| n-1\right\rangle _{1}\right) F\left( \left| 0\right\rangle
_{4}\text{ }\left| 1\right\rangle _{4}\cdots \left| n-1\right\rangle
_{4}\right) ^{t}
\end{eqnarray*}

where

\begin{equation}
\begin{array}{c}
\left| \psi _{14}\right\rangle =\left( \left| 0\right\rangle _{1}\text{ }%
\left| 1\right\rangle _{1}\cdots \left| n-1\right\rangle _{1}\right) F\left(
\left| 0\right\rangle _{4}\text{ }\left| 1\right\rangle _{4}\cdots \left|
n-1\right\rangle _{4}\right) ^{t} \\ 
F=A\overline{B}C/\rho .\left( \overline{B}_{ij}=B_{ij}^{\ast }\right) \\ 
\rho =\sqrt{tr\left\{ \left( A\overline{B}C\right) \left( A\overline{B}%
C\right) ^{\dagger }\right\} }.%
\end{array}
\label{aaa}
\end{equation}

Further more, there are $n$ bipartite entangled states $\left\{ \left| \phi
_{12}\right\rangle \left| \phi _{34}\right\rangle \cdots \left| \phi
_{\left( 2n-1\right) 2n}\right\rangle \right\} $ and $\left| \phi _{\left(
2k+1\right) \left( 2k+2\right) }\right\rangle =\mathbf{e}_{2k-1}\mathbf{A}%
_{k}\mathbf{e}_{2k}^{t}$ $\left| \phi _{\left( 2k\right) \left( 2k+1\right)
}^{^{\prime }}\right\rangle =\mathbf{e}_{2k}\mathbf{B}_{k}\mathbf{e}%
_{2k+1}^{t}$,then through $n-1$ quantum swappings the final entangled state
of particle $1$ and particle $2n$ is $\left| \psi _{1\left( 2n\right)
}\right\rangle =\mathbf{e}_{1}\mathbf{Fe}_{2n}^{t}$,where

\begin{equation}
\mathbf{F=}\left( \underset{k=1}{\overset{n-1}{\prod }}\mathbf{A}_{k}%
\overline{\mathbf{B}}_{k}\right) \mathbf{A}_{n}.
\end{equation}

In fact, quantum communication can be realized by sending one of the
entanglement pairs, while during long way transformation the entanglement
will decrease and some information will be losted. However we can overcome
this problem by a series quantum swapping as shown in the Fig $1$ (where $%
A_{k}$ means the $k$-th EPR pairs, and the $B_{k}$ is the $k$-th joint
measerment). The calculation of the series quantum swapping is very complex
even $n$ is a large number, but by the notation, the calculation become
concise.

\begin{equation*}
\FRAME{itbpF}{5.047in}{2.6922in}{0in}{}{}{seswap11new.jpg}{\special{language
"Scientific Word";type "GRAPHIC";maintain-aspect-ratio TRUE;display
"USEDEF";valid_file "F";width 5.047in;height 2.6922in;depth
0in;original-width 2.06in;original-height 1.094in;cropleft "0";croptop
"1";cropright "1";cropbottom "0";filename 'C:/Documents and
Settings/rechard/My Documents/lunwenpic/seswap11new.jpg';file-properties
"XNPEU";}}
\end{equation*}

\subsection{Bipartitie state teleportation}

Earlier studies have been confined to the teleportation of single-body
quantum states \cite{tel}, which is only a special case. Recently
teleportation of multi-body quantum states was attracted more \ attentions[$%
12$-$13$], the studies are only confined to teleportation of bipartite
entangled state. In this paper we study on how to teleport a bipartite
states by two EPR states based by the notation[$11$], and construct Bob's
local operation.

Supposed that \emph{Alice} was going to teleport a bipartite state $\left|
\phi \right\rangle _{12}=\underset{ij}{\Sigma }A_{ij}\left| i\right\rangle
_{1}\left| j\right\rangle _{2}=\overrightarrow{e}_{1}A\overrightarrow{e}%
_{2}^{t}$ to \emph{Bob}, she shared two known entanglement states $\left|
\alpha \right\rangle _{34}=\overrightarrow{e}_{3}B\overrightarrow{e}_{4}^{t}$
and $\left| \beta \right\rangle _{56}=\overrightarrow{e}_{5}C\overrightarrow{%
e}_{6}^{t}$ with \emph{Bob}. \emph{Alice} performed a joint states
measurement on particle $2$ and particle $3$, the result $\left| \delta
\right\rangle _{23}=\overrightarrow{e}_{2}D\overrightarrow{e}_{3}^{t}$, by
the eq.($1$) we get the state of particle $1$ and particle $4$ $\left|
\gamma \right\rangle _{14}=\overrightarrow{e}_{1}M\overrightarrow{e}_{4}^{t}=%
\overrightarrow{e}_{4}M^{t}\overrightarrow{e}_{1}^{t}$, where $M=\dfrac{A%
\overline{D}B}{\rho _{1}}$. \ Then \emph{Alice} made another joint states
measurement on particle $1$ and particle $5$, the result $\left| \eta
\right\rangle _{15}=\overrightarrow{e}_{1}F\overrightarrow{e}_{5}^{t}$, by
the eq.($1$) the state of particle $4$ and particle $6$ $\left| \varphi
\right\rangle _{46}=\overrightarrow{e}_{4}N\overrightarrow{e}_{6}^{t}$ was
easily obtained, where $N=\dfrac{M^{t}\overline{F}C}{\rho _{2}}=\dfrac{B^{t}%
\overline{D}^{t}A^{t}\overline{F}C}{\rho _{1}\rho _{2}}$.\qquad

\begin{equation*}
\left| \varphi \right\rangle _{46}=\left| \varphi \right\rangle _{64}=%
\overrightarrow{e}_{4}\dfrac{B^{t}\overline{D}^{t}A^{t}\overline{F}C}{\rho
_{1}\rho _{2}}\overrightarrow{e}_{6}^{t}=\overrightarrow{e}_{6}\dfrac{A%
\overline{D}B\overline{F}C}{\rho _{1}\rho _{2}}\overrightarrow{e}_{4}^{t}
\end{equation*}

\emph{Alice} told \emph{Bob} her joint states measurements, \emph{Bob}
perform a unitary operator $U_{4}$ on his particle $4$,

\begin{equation}
U_{4}\left| \varphi \right\rangle _{64}=\left| \phi \right\rangle _{64}\text{
where }U_{4}=\left( \dfrac{\overline{D}B\overline{F}C}{\rho _{1}\rho _{2}}%
\right) ^{-1}
\end{equation}%
,\emph{Bob} captured the state $\left| \phi \right\rangle $. We find that
Bob's local operator is independent with the state $\left| \phi
\right\rangle _{12}$.

\section{Conclusion}

After introduced a new notation for the quantum information, we obtained a
criteria for pure bipartite entangled state and a theorem for bipartites
quantum teleportation by two EPR pairs. We constructed the operation $U$
performed by the \emph{Bob} on one of his qubit in order to obtain perfect
replica of teleported states.

\bigskip


\vspace*{0.25cm}

\begin{thebibliography}{99}
\bibitem{1} A. Einstein, B. Podolosky and N. Rosen, Phys. Rev. \textbf{47}
777 (1935).

\bibitem{code} C. H. Bennett and S. J. Wiesner, Phys. Rev. Lett. \textbf{69}
,2881 (1992).

\bibitem{sw} M. Zukowski, A. Zeilinger, M. A. Horne, and A. K. Ekert,Phys.
Rev. Lett. 71, 4287 (1993)

\bibitem{tel} C. H. Bennett et al. Phys. Rev. Lett. \textbf{70} 1895 (1993).

\bibitem{ctel} L. Vaidman, Phys.Rev. A \textbf{49} 1473 (1994).

\bibitem{6} S. L. Braunstein and H. J. Kimble, Phys. Rev. Lett. \textbf{80}
869 (1998).

\bibitem{7} S. Bose and V. Vedral, Phys. Rev. A \textbf{61,} 040101(2000).

\bibitem{8} S. L. Braunstein, G. M. D'Ariano, G.L. Milburn and M. F. Sacchi,
Phys.Rev.Lett. \textbf{84 } 3486 (2000).

\bibitem{9} L. Accardi and M. Ohya, Quantum information Proceedings of the
First International Conference Meijo University, Japan 4 - 8 November 1997

\bibitem{10} L. P. Hughston, R.J. Ozsa and W.K. Wootters, Phys. Lett. A 
\textbf{183} 14 (1993)

\bibitem{11} C.Li,H.S.Song and Y.X.Luo,Phys. Lett. \textbf{A 297} (2002) 121

\bibitem{12} J.Lee and M.S. Kim,Phys. Rev.Lett.\textbf{84} (2000)4236

\bibitem{13} X.Wang, Phys. Rev.\textbf{A64} (2001)022302

\bibitem{14} J. Lee, H. Min and S.D. Oh, Phys.Rev. A \textbf{66} (2002)052318

\bigskip
\end{thebibliography}
\end{document}